# Benchmarking of the construct of dimensionless correlations regarding batch bubble columns with suspended solids:
## Performance of the *PRESSURE TRANFORM* approach

Jordan Hristov

**Abstract** – *Benchmark of dimensionless data correlations pertinent to batch bubble columns (BC) with suspended solids has been performed by the pressure transform approach (PTA). The main efforts have addressed the correct definition of dimensionless groups referring to the fact that solids dynamics and the bubble dynamics have different velocity and length scales. The correct definition of the initial set of variable in the classical dimensional analysis depends mainly on the experience of the investigator while the pressure transform approach (PTA) avoids errors at this initial stage. PTA addresses the physics of the phenomena occurring in complex systems involving many phases and allows straightforward definitions of dimensionless numbers.*

*Keywords*: bubble column, pressure transform, dimensionless correlation construct

## I. Introduction

Bubble columns (BC) are widely encountered in the process industry for performance mass transfer operations with or without suspended solids [1]-[3]. Moreover, such devices are objects of many studies continuously appearing in the literature. The present benchmark exercise addresses data correlations in bubble columns when many process variables are involved. The main problem in transferring and dissemination of knowledge in this area of the process hydrodynamics is the lack of unified approach in data correlations in contrast to the fields of heat and mass transfer [4],[5]: the book of Fan [1] and many articles [5]-[8] following the modern prolific style of scientific publications are good examples.

Many correlations concerning bubble columns [3], [6], [7] are not entirely dimensionless; contain dimensional terms and coefficients that actually avoid transferring the results from one study to others. The classical gas-holdup equation [2]

$$\varepsilon_G = aU_G^n \qquad (1)$$

is a good example of such incorrect data scaling irrespective of its large use in practice. In general, the origin of this problem is due to two basic reasons:
a) Formalistic creation of equations using other already published relationships without any critical examinations when applying to new data and process conditions.
b) Incorrect performance of dimensional analysis in its classic "blind approach" using the $\pi - theorem$ [8],[9].

The benchmark developed in this work shows how errors in creation of dimensionless groups may be avoided by application of the "pressure transform" approach (PTA) [10]-[12]. This method address the physical basis in the dimensionless group formation and avoids errors in the formation of the initial set of process variables, thus avoiding the basic source of error in application of the $\pi - theorem$. All these general standpoints will be exemplified in the next sections.

## II. Pressure transforms approach
### *General suggestions*

The pressure transform method was developed for complex systems involving various physical fields [10] - [12] and mainly applied to fluidized granular systems. When several physical fields are involved in the process, then PTA presents their action on a certain control volume as surface forces having dimensions of $N/m^2$ (the dimension of pressure). In this context, let us consider the common example of fluid flow through a control volume in Cartesian co-ordinates (see Fig. 1a). The force balance over the control volume leads to the Navier equations expressed in terms having dimensions $N/m^2$. The transition to various models depends on the rheology of the medium that in the simple case of Newton aw $\tau = -\mu \dfrac{\partial U}{\partial y}$ leads we get the Navier - Stockes equations [4],[5]

$$\tau_{xy} = \tau_{yx} = \mu\left(\frac{\partial U_x}{\partial y} + \frac{\partial U_y}{\partial x}\right) \qquad (2b)$$

and

$$\tau_{xx} = 2\mu\frac{\partial U_x}{\partial x} - \frac{2}{3}\mu\left(\frac{\partial U_x}{\partial x} + \frac{\partial U_y}{\partial y}\right) \qquad (2c)$$

According to Fig. 1b, the ratio of the dynamic pressure $P_U = \rho_f U_f^2$ in the y-direction to the that produced by the gravity $\rho_f g l$ (i.e. produced by the body forces), where is the length scale, yield the ratio

$$\frac{P_U}{P_g} = \frac{\rho_f U_f^2}{\rho_f g l} = \frac{U_f^2}{g l} = Fr \qquad (3a)$$

that is the Froude number.

Similarly, the ratio of the dynamic pressure along the x-direction to the viscose stresses

$$P_\mu = \frac{\mu\left(\frac{U_f}{l}\right)}{\rho_f U_f^2} = \frac{\mu}{\rho U_f l} = \frac{1}{Re} \qquad (3b)$$

Hence, the Reynolds number was just defined.

The common approach to define dimensionless numbers when a deterministic model is available is to create ratios of terms in (2a) after adimensialization. In fact, this procedure simply means formation of ratios of specific pressure (surface forces) acting on the control volume surface; this matches the idea of PTA-see Fig.1.c. The following benchmark example will show how PTA works in the complex case of BC dynamics where is hard to create deterministic models, or at least is hard to handled and use to create dimensionless correlations

## III. Bubble columns with suspended solids- batch mode

*III.1 Initial sets of media and process variables*

Consider a bubble column (see Fig. 2a) consisting of vessel with diameter $D$ and filled by a liquid with a depth $L$ (between the gas input and the top liquid level). The first step of PTA is to define the media (the phase) involved in the process and the process variables in accordance with the template given by Table I, namely:

1) **Extensive variables** depending on the mass of the media and including the transport coefficients (the dynamic viscosity, for example);
2) **Intensive variables** which characterize the forces acting on the media in the column. Each medium has its own variables related either to group 1 or group 2.

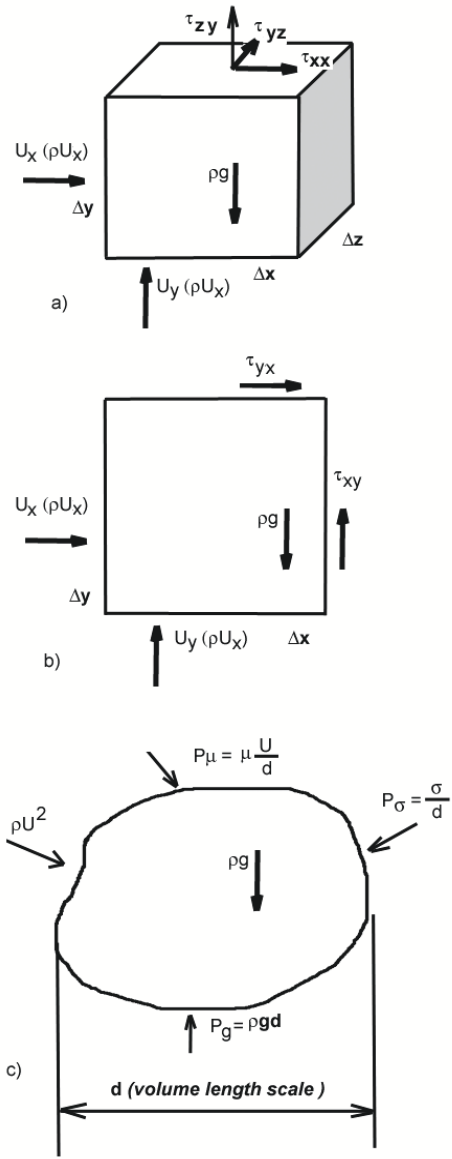

Fig.1 Control volume of a flowing medium
a) Classical 3-D control volume in Cartesian coordinate used to derive the Navie stokes equations.
b) Classical 2-D control volume in Cartesian coordinate used to derive the Navie stokes equations in a plane flow
c) General control volume showing how all physical fields involved in the process generate surface forces.

For seek of simplicity, the basic principle of PTA will be illustrated with the 2-D control volume in Fig.1b. In this case the force balance along the x-axis results in

$$\rho\frac{\partial U}{\partial t} + \frac{\partial(\rho U_x^2)}{\partial x} + \frac{\partial(\rho U_x U_y)}{\partial y} = \frac{\partial p}{\partial x} + \frac{\partial \tau_{xx}}{\partial x} + \frac{\partial \tau_{xy}}{\partial y} \qquad (2a)$$

with

The vessel is represented by the gravity pressure produced at the column bottom and implicitly by the gas superficial velocity defined through the column diameter $D$. Commonly, the variables summarized in Table I are treated by the classical dimensional analysis (DA) providing dimensionless groups requiring further re-arrangements and analysis.

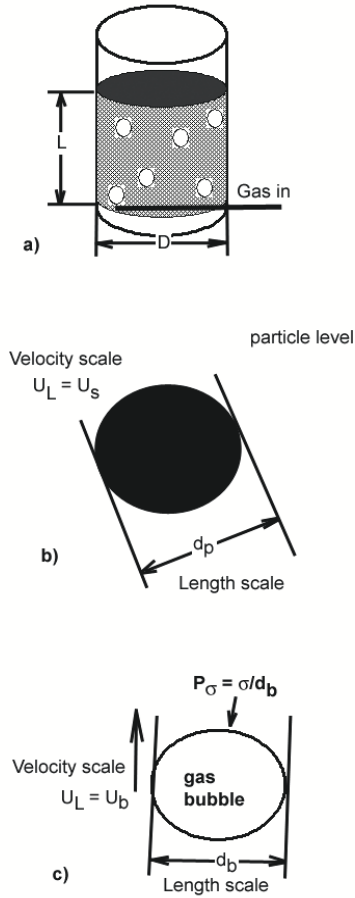

Fig. 2. Bubble column problem at different levels
a) Macroscopic scales defined by the column geometry
b) Flow problem and scales around a suspended particle
c) Flow problem and scales around a rising gas bubble

TABLE I
DEFINITION OF THE MEDIA PARTICIPATING
THE PROCESS AND BASIC PROCESS VARIABLES

|  | Dimensional variables | |
|---|---|---|
| Medium | Extensive variables | Intensive variables |
| Liquid | $\rho_L$, $\mu_L$, $d_p$, $g$ | $U_L$ |
| Solids | $\rho_s$, $d_p$, $g$, $C_W$ | $U_S$ |
| Bubbles | $\rho_g$, $\rho_L$, $\sigma_L$, $d_b$ | $U_b$ |
| Gas | $\rho_G$, $\mu_G$, $g$, | $U_G$ |
| Vessel | $\rho_L$, $g$, $L$ | |

*III.2. Reduction in the number of media involved and the process variables*

The second step, related to the PTA approach avoids the initial step of DA and refers to the process physics. More precisely, let us arrange the initial variables as it is done in the first two columns of Table II-A transforming them into new ones in Table II-B. First of all, taking into account that in liquid-solids systems the buoyancy takes place a complex medium termed "liquid –solids" replaces the initial ones "liquid" and "solids". Similarly, since the gas flow at the vessel bottom is responsible only to the energy input into BC a "new medium "gas-vessel" represents both the gas flow and the vessel geometry. Each new medium is characterized by its *extensive* an *intensive* variables combining the old ones. This step expels from the group of the intensive variable the solids settling velocity $U_S$ because, in fact, it is equal to $U_L$ - the relative solid-liquid velocity at the particle level scale (see the comments below and Fig. 2b).

*III.3. Surface forces (pressures) : definition*

The principle step of PTA is to define the surface forces (pressures) produced by each physical field acting on the media. In the BC case, we have
- **Fluid dynamic pressure**,
- **Gravity induced pressure,** representing the body forces ,
- **Viscous drag pressure (**stress**)**
- **Surface tension pressure**

In accordance with these general definitions, *each medium has its own velocity and length scales* closely related to the process physics. In this context, for example, the particle motion in the liquid depends on the viscous drag forces and the particle-liquid relative velocity $U_L$, while the length scale is defined by the particle diameter $d_p$ -see Fig.2b. At the same time the second disperse phase, the bubbles (see Fig.2c) have its own velocity scale defined by the *bubble rise velocity* $U_b$ and a length scale defined by the bubble diameter $d_b$. With these *basic scale*s, the pressures produced by the physical fields form by the initial set of *process variables* as it is shown in Table II-A. The *set of basic pressures* after the reduction of the member of participating media and process variables are shown in Table II-B.

*III.4. Pressure scale invariant to process conditions and formation of dimensionless ratios*

The forth step of PTA is to define the *invariant pressure scale*. *The invariant scale* should be known *prior to start the experiments* and *independent of its* *performance*. Therefore, the only scale satisfying these conditions is conditions is $P_g$ because the gravity does not vary during the experiments and each $P_g$ could be calculated (defined) at the beginning. As to the "liquid-solids" medium this approach results in two known dimensionless numbers: Froude number $Fr_L$ and Reynolds number $Re_L$. Similar numbers appear in the cells of Table II-C corresponding to the other media involved in the process: $Fr_b$, $Re_b$ and the Weber number, $W_b = (P_{\sigma-b}/P_{U-b})$, all of them relevant to the gas bubble motion.

TABLE II-A
DEFINITION OF THE PRESSURES ASSERTED BY THE PHYSICAL FIELDS INTERACTING THE PROCESS WITHOUT REDUCTION OF THE MEDIA INVOLVED

| Medium | Dimensional variables | | Pressures due to physical effects | | | |
|---|---|---|---|---|---|---|
| | Extensive variables | Intensive variables | Fluid dynamic pressure | Gravity induced pressure | Viscous drag pressure | Surface tension pressure |
| Liquid | $\rho_L, \mu_L, d_p, g$ | $U_L$ | $P_{U-L} = \rho_L U_L^2$ | $P_{g-L} = \rho_L g d_p$ | $P_{\mu-L} = \mu_L (U_L/L)$ | |
| Solids | $\rho_s, d_p, g, C_W$ | $U_S$ | $P_{U-S} = \rho_L U_S^2$ | $P_{g-S} = \rho_L g d_p$ | $P_{\mu-S} = \mu_L (U_L/d_p)$ | |
| Bubbles | $\rho_G, \rho_L, \sigma_L, d_b$ | $U_b$ | $P_{U-b} = \rho_L U_b^2$ | $P_{g-b} = (\rho_L - \rho_g) g d_b$ $P_{g-b} \approx \rho_L g d_b$ | $P_{\mu-b} = \mu_L (U_b/d_b)$ | $P_{\sigma-L} = \dfrac{\sigma_L}{d_b}$ |
| Gas | $\rho_G, \mu_G, g$ | $U_G$ | $P_{U-G} = \rho_G U_G^2$ | | | |
| Vessel | $\rho_L, g, L$ | | | $P_{H-L} = \rho_L g L$ | | |

TABLE II-B
FORMATION OF THE PRESSURE RATIOS TAKING INTO ACCOUNT THE REDUCTION IN THE MEDIA INVOLVED

| Medium | Dimensional variables | | Pressures due to physical effects | | | |
|---|---|---|---|---|---|---|
| | Extensive variables | Intensive variables | Fluid dynamic pressure | Gravity induced pressure | Viscous drag pressure | Surface tension pressure |
| Liquid-Solids | $(\rho_S - \rho_L), \mu_L, d_p, g, C_W$ | $U_L$ | $P_{U-L} = (\rho_S - \rho_L)$ | $P_{g-L} = (\rho_S - \rho_L) g d_p$ | $P_{\mu-L} = \mu_L (U_L/L)$ | |
| Gas -Vessel | $\rho_G, \rho_L, \mu_G, g, L$ | $U_G$ | $P_{U-G} = \rho_G U_G^2$ | $P_{H-L} = \rho_L g L$ | | |
| Bubbles | $\rho_g, \rho_L, \sigma_L, d_b$ | $U_b$ | $P_{U-b} = \rho_L U_b^2$ | $P_{g-b} = (\rho_L - \rho_g) g d_b$ $P_{g-b} \approx \rho_L g d_b$ | $P_{\mu-b} = \mu_L (U_b/d_b)$ | $P_{\sigma-L} = \dfrac{\sigma_L}{d_b}$ |

TABLE II-C
DIMENSIONLESS NUMBERS DEFIND BY THE RATIOS OF THE CHARACTERISTIC PRESSURES

| Medium | Dimensional variables | | Pressures ratios (dimensionless) | | | |
|---|---|---|---|---|---|---|
| | Extensive variables | Intensive variables | Fluid dynamic pressure | Gravity induced pressure | Viscous drag pressure | Surface tension pressure |
| Liquid-Solids | $(\rho_S - \rho_L), \mu_L, d_p, g, C_W$ | $U_L$ | $\dfrac{P_{U-L}}{P_{g-L}} =$ $\dfrac{U_L^2}{g d_p} = Fr_L$ | - | $\dfrac{P_{U-L}}{P_{\mu-L}} =$ $\dfrac{(\rho_S - \rho_L) U_L^2}{\mu_L (U_L/L)} = \mathrm{Re}_L$ | |
| Gas -Vessel | $\rho_G, \mu_G, g$ | $U_G$ | $P_{U-G} = \rho_G U_G^2$ | $\dfrac{P_{U-G}}{P_{H-L}} = \dfrac{\rho_G U_G^2}{\rho_L g L}$ | | |
| Bubbles | $\rho_G, \rho_L, \sigma_L, d_b$ | $U_b$ | $\dfrac{P_{U-b}}{P_{g-b}} =$ $= \dfrac{U_b^2}{g d_b} = Fr_b$ | | $\dfrac{P_{U-b}}{P_{\mu-b}} =$ $= \dfrac{\rho_L U_b^2}{\mu_L (U_b/d_b)} = Re_b$ | $\dfrac{P_{\sigma-b}}{P_{U-b}} =$ $= \dfrac{(\sigma_L/d_b)}{\rho_L U_b^2} = We$ |

TABLE II-D
REDUCTION OF THE SET OF DIMENSIONLESS NUMBERS ELIMINATING $U_L$ AND $U_b$ FROM THEM

| Medium | Dimensional variables | | Dimensionless variables | |
|---|---|---|---|---|
| | Extensive variables | Intensive variables | Initial set | Reduced (final) set |
| Liquid-Solid | $(\rho_S - \rho_L)$, $\mu_L$, $d_p$, $g$, $C_W$ | $U_L$ | $Fr_{L-S}$, $\mathrm{Re}_{L-S}$, $\dfrac{(\rho_S - \rho_L)}{\rho_L}$ <br> Reduction : <br> $\dfrac{(\mathrm{Re}_{L-S})^2}{Fr_{L-S}} = Ga_{L-S}, \dfrac{\rho_s - \rho_L}{\rho_L}$; <br> $Ar = Ga_{L-S}\left(\dfrac{\rho_s - \rho_L}{\rho_L}\right)$, $C_W$ | $Ar$, $C_W$ |
| Gas - Vessel | $\rho_G$, $\mu_G$, $g$ | $U_G$ | $Y = \dfrac{\rho_G U_G^2}{\rho_L g L}$ | $Y = \dfrac{\rho_G U_G^2}{\rho_L g L}$ |
| Bubbles | $\rho_G$, $\rho_L$, $\sigma_L$, $d_b$ | $U_b$ | $\dfrac{U_b^2}{g d_b} = Fr_b$ ; $\dfrac{\rho_L U_b d_b}{\mu_L} = \mathrm{Re}_b$ ; $\dfrac{(\sigma_L/d_b)}{\rho_L U_b^2} = We$ <br> $\dfrac{(\mathrm{Re}_b)^2}{Fr_b} = \dfrac{d_b^3 \rho_L g}{\mu_L^2} = Ga_b$ <br> $F_{r-b} We = \dfrac{\sigma_L/d_b}{\rho_L g d_b} = Bo$ | $\dfrac{d_b^3 \rho_L g}{\mu_L^2} = Ga_b$ <br> $\dfrac{P_{\sigma-L}}{P_{g-b}} = \dfrac{(\sigma_L/d_b)}{\rho_L g d_b} = Bo$ |

The main problem appearing at this stage of the PTA comes from the velocity scales $U_L$ and $U_b$ *which are unknown and cannot be defined prior to the experiments* even though *they exactly represent the process physics*. This *trivial problem in the fluid mechanics* can be simply solved by the ratios $(Re_L)^2/Fr_L$ and $(Re_b)^2/Fr_b$ eliminating the unknown velocity $U_L$ and $U_b$. These ratios are Archimedes, $Ar$ and Galileo $Ga_b$ numbers for the "liquid-solids" and the bubbles, respectively: see the corresponding cells in Table II-D.

*III. 5. Initial set of dimensionless variables provided by PTA*

The approach applied in III.4 reduces the numbers of dimensionless numbers but the bubble rise velocity $U_b$ still remains in the Weber number. Here, the process of determination of dimensionless numbers can stop if the bubble size is quite enough to assume the bubble rise velocity as a constant value. For bubble columns with liquid having properties close to those of water, $U_b$ is almost constant if the bubble size is beyond $1\,mm$. With these assumption the dimensionless groups involved in BC dynamics are (see Table II-D):

- **Independent variables** : $Ar, C_W, Ga_b, Wb$
- **Dependent variable:** depends on the process parameter we try to define as critical one. These are exemplified below.

*III. 6. Reduced set of dimensionless variables*

However, if the bubbles rise velocity is not constant [13], [14], then the product (see Table II-D) $F_{r-b} We = \dfrac{\sigma_L/d_b}{\rho_L g d_b} = Bo$ defines the gas bubble Bond number. Hence, the independent dimensionless variables now are: $Ar, C_W, Ga_b, Bo$

# IV. Constructs of possible dimensionless correlations

First of all we have to mention that the target of the any experiment performed defines the principle independent variable of the process and those from the group of the independent variables. Several basic constructs of dimensionless correlations relevant to batch bubble columns with suspended solids will be discussed, among them.

*IV.1. Critical velocity*
*(Minimum Suspension point, for instance)*

The critical gas velocity $U_G$ corresponding the "off-bottom point" of complete solids suspension is **the dependent process variable**. The adimensialization of $U_G$ is $Y = \dfrac{\rho_G U_G^2}{\rho_L g L}$ because $U_G$ only characterizes the energy input to the column that is required to suspend the particles and to balance the hydrostatic liquid pressure. Hence, we have

Hence, the basic form of the relationship is:

$$\frac{P_{U-G}}{P_{H-L}} = \frac{\rho_G U_G^2}{\rho_L g L} = Y = f_U[Ar, We, C_W] \Rightarrow$$

$$\Rightarrow U_G = \sqrt{\frac{\rho_L g L}{\rho_G}} f_U[Ar, We, C_W] \quad (4)$$

*IV.2. Gas holdup*

The gas holdup $\varepsilon_G$ is dimensionless and it immediately can be added to the group of the final set as a **dependent variable**. All other numbers (the dimensionless energy input due to the gas flow, too) form the group of the **independent variable**s. Thus, the basic relationship is

$$\varepsilon_G = f_\varepsilon[Ar, We, C_W, Y] \Rightarrow f_\varepsilon[Ar, We, C_W, Y] \quad (5)$$

*IV.3. Pressure Drop*

In general, the pressure drop across the column is a *dependent process variable* depending the gas flow and other process parameters. Hence, the pressure drop with respect to the blowing gas has to be introduced in the initial set of dimensional variables corresponding to the *Gas* or *Gas-vessel* group. Consequently, applying all steps of PTA from Table II-A to Table II-B we get the ratio $\Delta P / \rho_L g L$ as dependent dimensionless process variables. The ratio $\Delta P / \rho_L g L$ has an order of magnitude of unity. Therefore, the general expression in a dimensionless form should be

$$\frac{\Delta P}{\rho_L g L} = f_p[Ar, Ga_b, Bo, C_W, Y] \quad (6)$$

# V. Power-law functions and pre-factors

According to the rules of scaling [15], [16], the **dependent** and the **independent** dimensionless variable should be related through a power-law function. Because *We* and *Ar* are preliminarily known as representative of the operating and initial conditions, respectively, *they contribute the power-low only as pre-factors*. These dimensionless variables become **parameters** and the correlations should be expressed as

$$(dependent\,variable) = k(C_W)^m (Ar, We) \quad (7)$$

Hence, in the cases discussed above we have some typical situations, among them:

*V.1.Critical velocity*
*(Minimum Suspension point, for instance)*

$$Y = k_Y (C_W)^m \varphi_U[Ar, We] \quad (8)$$

The numerical pre-factor $k_y$ and the exponent *m* have to be obtained by fitting to experimental data.

*V.2. Gas holdup*

Now, all dimensionless variables, **except** $Y \sim U_G^2$ **are parameters**. Hence, the expression should be

$$\varepsilon_G = k_\varepsilon Y^E \varphi_\varepsilon[Ar, Ga_b, Bo, C_W] \quad (9)$$

The numerical pre-factor $k_\varepsilon$ and the exponent *E* have to be obtained by fitting to experimental data.

*V.3. Pressure Drop*

This case resembles that concerning the pressure drop correlations. Hence, the independent variable is $Y \sim U_G^2$, while the dependent is $Z_p = \Delta P/(\rho_L g L)$, namely

$$Z_p = k_p Y^q \varphi_p[Ar, Ga_b, Bo, C_W] \quad (10)$$

The numerical pre-factor $k_p$ and the exponent *q* have to be obtained by fitting to experimental data.

## VI. Discussion

The "pressure transform" approach developed in this work and exemplified by the batch bubble column case clearly demonstrates how the process physics affects the definitions of either or the process variable or the scales (velocity and length) involved in each particular sub-process. This initial comment addresses some basic points in the PTA technology that have to be clarified.

The analysis has employed velocity scales $U_L$ and $U_b$ which refer exactly to the physics even though they are unknown prior to the experiments and consequently eliminated as the number of dimensionless numbers was refined. In contrast, the superficial gas velocity $U_G$ was not included in the basic process variables and did not contribute to the dimensionless numbers developed. However, if the gas superficial velocity $U_G$ is defined as a unique representative of the family of velocities controlling the process and a classical dimensional analysis (DA), for instance, is performed, then the outcome will be $Re$ or $Fr$ numbers both to the solids and the bubbles having common velocity scale $U_G$. Nevertheless, $U_G$ **is not involved** in either the solids suspension (depending on $U_L$) or the bubble motion (controlled by $U_b$), even though the algebraic procedure of DA is performed correctly.

The error could be fixed, to some extent, by formation of either particle Archimedes number $Ar$ or the bubble Galileo number $Ga_b$. The main indicator that something was wrong in the formation of the dimensionless numbers is the appearance of the gas superficial velocity in the Weber number $We = \dfrac{(\sigma_L/d_b)}{\rho_L U_G^2}$. This form of *We* generally contradicts its definition as ratio of the *surface tension pressure* $\sigma_L/d_p$ to the *dynamic pressure* $\rho_L U_b^2$ of **a rising gas bubble**. The bubble rise velocity $U_b$ plays the same role at $U_L$ with respect to the solids in the liquid. However, all these sub-processes **do not involve the gas superficial velocity** $U_G$ and its use in definitions of the above mentioned dimensionless numbers is incorrect. However, we have to correct to the reader because other standpoints have to be mentioned. The common form of the Weber number in case of perforated spargers (plates or tubes) is [17]:

$$We = \frac{\rho_G U_{G,0} d_0}{\sigma} \quad (11)$$

where $U_{G,0}$ is the gas superficial velocity at the sparger orifice of diameter $d_0$.

This equation (11) is valid only in the case of *bubble formation at the orifice* and <u>cannot be used when the bubble rises through the liquid</u>. To some extent, if the deep physical analysis is applied again, the medium "gas-vessel" should incorporates as variables $U_{G,0}$ and $d_0$ but we have to forget about the bubble rise action due to emerging ambiguity in definition of the Weber number appearing immediately : one *We* is defined for the rising bubbles and other one for the bubble formation (eq. 11). The bubble rise promotes liquid mixing and consequent suspension of the solids while the bubble formation at the sparger orifice is not related to this phenomenon.

The bubble diameter also varies along the column radius [18]-[20] and depends on the liquid rheology [21]. Hence, different approaches are possible but if an average bubble diameter is used as a length scale, the problem could be solved to some extent: the dimensionless numbers involving $d_b$ could be correctly calculated. The Sauter diameter $d_{32}$, for example is a good solution, namely [22]

$$d_{32} = \frac{\sum_{1}^{N} d_i^3}{\sum_{i-1}^{N} d_i^2} \quad (12)$$

This implies collection of a lot of experimental data about $d_b$ and a clear standpoint: does such average value represent adequately the process at issue? In this context, Krishna et al. [23] show that $U_b \propto d_b$, so variations in bubble diameter results in variations of $U_b$ a variable Weber number. From this standpoint, the Weber number, to some extent, is a dependent process variable since both $U_b$ and $d_b$ vary. However, upon certain experimental conditions the Weber number oscillates around a certain mean value, and for practical use it could be considered as a process parameter rather then as dependent variable.

The comments in the preceding paragraphs strongly address the advantages of the PTA with respect to the classical DA, among them:

1) The wrong definition of the process variable could be done and some media parameters could be forgotten if the classical dimensional analysis is applied in a "blind manner ".

2) The definition of surface forces (pressures) produced by each physical field involved avoids either the pressures to be defined in a wrong way or some process parameters to be forgotten.

3) As a positive outcome at the intermediate stages of PTA, each process variable and media parameter are *properly immobilized* by respective *pressure* (relevant to a given physical field) and consequently by appropriate dimensionless numbers.

4) Therefore, the algebraic formality of the classic DA is replaced by a simple approach requiring the process physics to be known even though deterministic models cannot be defined and handled.

5) The simple templates of correlations suggested above show how the process variables in the specific case of Bubble Column hydrodynamics should be involved in data correlations.

6) PTA technology needs only a good knowledge and tables such as those developed in the present article. PTA does not need algebraic manipulations in contrast to the classic DA.

Last but not least, the author thanks to many authors publishing works on bubble columns and providing experimental data correlated by dominating incorrectly designed relationships. This was the reason to start to work on PTA application to bubble columns. Moreover, the present article, and particularly the research performed were provoked by many manuscripts passing through my hands due to editorial and reviewer duties. Hence, anything, even wrong results, may provoke studies leading to better understanding of certain problems at issue. Creation of equations relating experimental data and dimensionless equations is an art requiring deep understanding of the physics of the modelled processes that with a good knowledge becomes a science.

## VI. CONCLUSIONS

The present work addresses application of the pressure transform approach to the constructs of dimensionless relationships pertinent to bubble column dynamics. It is an instructive article showing all steps and potential pitfalls of the PTA technology. The main outcomes of the analysis performed may be outlined briefly as:
- The PTA approach avoids errors in the formation of the initial set of process variable and consequent algebraic mistakes in application of the classical dimensionless analysis.
- The formation of a set of dimensionless process variables with homogeneous dimensions of $N/m^2$ allows easily formation of dimensionless numbers as ratios of the characteristic pressures. These, in fact, are ratios of the fluxes of the physical fields acting on the system of interest and entering the control volume.
- The analysis performed allows the correct constructs of dimensionless correlations relevant to BC to be designed. This approach clearly defines the dependent and independent process variables and the forms of the power-law relationships.

**Nomenclature**

$C_W$ - dimensionless concentration of solids: with respect to the liquid, $\left(\dfrac{kg\ solids}{kg\ liquid}\right)$ or $\left(\dfrac{volume\ of\ solids}{volume\ of\ the\ slurry}\right)$

$D$ -column diameter, $m$
$d$ -length scale (see Fig.1c and Fig. 2), $m$
$d_p$ -particle diameter, $m$
$d_b$ -bubble diameter, $m$
$f_U, f_\varepsilon, f_p$ -symbols denoting in general the dimensionless relationships (Eqs. 4, 5 and 6)
$k$ -dimensionless pre-factor in Eq.7.
$k_Y, k_\varepsilon, k_p$ - dimensionless pre-factors in Eq.8, 9 and 10.
$E$ - dimensionless exponent in Eq.9.
$L$ -initial liquid depth, $m$
$l$ -characteristic length scale (see Eqs 3a, b), $m$
$m$ -dimensionless exponent (Eqs. 7 and 8)
$p$ -pressure, $N.m^{-2}$
$P$ -pressure, $N.m^{-2}$
$q$ - dimensionless exponent in Eq.10.
$U$ -velocity (see the context), $m.s^{-1}$

*Greek letters*
$\varepsilon$ -gas holdup, (-)
$\mu$ -fluid dynamic viscosity, $(N.m^{-2}).s$
$\rho$ -density, $kg.m^{-3}$
$\tau$ - shear stress, $N.m^{-2}$
$\varphi_U, \varphi_p, \varphi_\varepsilon$ -general forms of the functions relating the process parameters in the power-law functions (see Eqs. 8, 9 and 10).

Subscripts
$f$ -fluid
$G$ -gas
$g$ -gravity produced
$L$ -liquid
$s$ -solids or settling
$U$ -fluid dynamic induced
$x$ - component along the x co-ordinate
$y$ - component along the y co-ordinate

$\mu$ -viscous
$\sigma$ -surface tension related

# References

[1] L.S. Fan, *Gas-Liquid-Solid Fluidization Engineering* (Butterworths, London, 1989).

[2] W.D.:Deckwer, *Bubble Column Reactor* (John Wiley and Sons, Chichester, 1992)

[3] R. Krishna, J.Ellenberger, C. Maretto C. Flow regime transition in bubble columns, *Int. Comm. Heat Mass Transfer*, **26** *(4) (1999) 467-475.*

[4] A.Bejan, Convection Heat Transfer (Wiley-Interscience Publication, New York, 1984).

[5] F.P. Incropera, D.P., DeWitt,. *Fundamentals of heat and mass transfer* (John Wiley & Sons, New York, 1996)

[6] H. Hikita,, S.Asai,, K.Tanigawa., K.Segawa, M.Kitao, Gas Hold-Up in Bubble Columns, *Chem. Eng. J.*, **20** *(1980) 59-67.*

[7] G.A. Hughmark,, Holdup and Mass Transfer in Bubble Columns", *Ind. Eng. Chem. Proc. Des. Dev.*, **6** *(1967) 218-221.*

[8] J.M.Kay, R.M. Nedderman,. *Fluid mechanics and transfer processes* (Cambridge University Press, 1990).

[9] D.P. Kessler, R.A.Greenkorn, Momentum*, Heat, and Mass Transfer* (Marcel Dekker, New York, 1999).

[10] J.Hristov, Magnetic field assisted fluidization-A unified approach. Part5. A Hydrodynamic Treatise on Liquid-solid fluidized beds**, *Reviews in Chemical Engineering*, **22** (4-5) *(2006) 195-375.*

[11] J. Hristov, Magnetic Field Assisted Fluidization: Dimensional analysis addressing the physical basis, *China Particuology*, **5***(1) (2007) 103-110.*

[12] J. Hristov, Magnetically Assisted Gas–Solid Fluidization in a Tapered Vessel: First Report with Observations and Dimensional Analysis, *The Canadian Journal of Chemical Engineering,* **86** *(2008), 3, 470-492* .

[13] R.M.Davies, F.I.Taylor, The mechanics of large bubbles rising through extended liquids and through liquids in tubes. *Proc. R. Soc. Lond.*, A 200 (1950) 375-390.

[14] Y. Zhang, J.A. Finch, A note on single bubble motion in surfactant solutions. *J. Fluid. Mech.* **429** *(2001) 63-66.*

[15] J. Palacios, *Dimensional Analysis* (Mc Millan, London, 1964).

[16] H.L. Langhaar, *Dimensional Analysis and Theory of Models* (John Wiley & Sons: New York, 1951).

[17[ A. Mersmann, Design and Scale up of the Bubble and Spray Columns, *Ger. Chem. Eng,* 1 *(1979) 1-11.*

18] G.M.Evans, E. Doroodchi, G.L. Lane, P.T.L.Koh, M.P.Schwartz, Mixing and gas dispersion in mineral flotation cells , *Chemical Engineering Research and Design*, **86** *(2008) 1350-1362.*

[19] D. Sikorski, H.Tabuteau, J.R.de Bruyn, Motion and shape of bubbles rising through a yield- stress fluid , *Journal of Non-newtonian Fluid Mechanics*, **159** *(2009) 10-16.*

[20] L.A.Mondy, C. Betallack, K.Thompson, J. Barney, A. Grillet, On bubbles rising through suspension of solid particles, *AIChE J.* **54 (4)** *(2008) 862-871.*

[21] R.P. Chhabra, *Bubbles, drops and particles in non-Newtonian fluids* (CRC Press, Boca Raton, Fla., 1993).

[22] R.Pohorecki, W. Moniuk, A. Zdrojkowski, P. Bielski, Hydrodynamics of Pilot Plant Bubble Column under Elevated Temperature and Pressure, *Chem. Eng. Sci.*, **56** *(2001)11-67.*

[23] R.Krishna, J.M. van Baten, M.I. Ureseanuy, J. Ellenberger, A scaleup strategy of bubble column slurry reactors, *Catalyst Today*, **66** *(2001) 199-207*.

## Authors' information

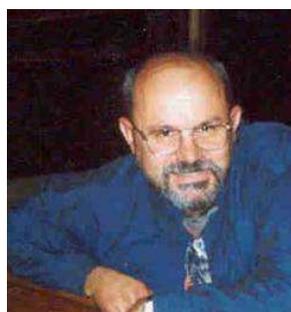

**Jordan Hristov** is associate professor of Chemical Engineering at the University of Chemical Technology and Metallurgy, Sofia, Bulgaria. He was graduated in 1979 as Electrical Engineer (MS equivalent) at the Technical University, Sofia, Bulgaria. His PhD thesis on the magnetically assisted fluidization was awarded by the University of Chemical Technology and Metallurgy in 1995. A/Prof. Hristov's research interests cover the areas of particulate solids mechanics, fluidisation, heat and mass transfer with special emphasis on scaling and approximate solution. The main branch of his research is devoted to magnetic field effects of fluidisation. Additionally, specific heat transfer topics are at issue, especially to thermal effects in accidents (fire). Relevant information is available at http://hristov.com/jordan .
.